\documentclass[prl,numerical,preprint,superscriptaddress]{revtex4-1}

\usepackage{amsmath} 
\usepackage[applemac]{inputenc}
\usepackage{graphicx} 
\usepackage{epstopdf}
\usepackage{textcomp}
\usepackage{hyperref}

\begin{document}

\title{A tank-circuit for ultrafast single particle detection in micropores}

\author{Abhishek Bhat}
\affiliation{Department of Materials Science and Engineering, University of Wisconsin-Madison, 1509 University Avenue, Madison, WI 53706, USA}
\author{Paul V. Gwozdz}
\affiliation{Center for Hybrid Nanostructures, Institute for Nanostructure and Solid State Physics, Universit\"at Hamburg, Luruper Chaussee 149, 22761 Hamburg, Germany}
\author{Arjun Seshadri}
\affiliation{Department of Electrical and Computer Engineering, University of Wisconsin-Madison, Madison, USA Madison, WI 53706, USA}
\author{Marcel Hoeft}
\affiliation{Center for Hybrid Nanostructures, Institute for Nanostructure and Solid State Physics, Universit\"at Hamburg, Luruper Chaussee 149, 22761 Hamburg, Germany}
\author{Robert H. Blick}
\affiliation{Department of Materials Science and Engineering, University of Wisconsin-Madison, 1509 University Avenue, Madison, WI 53706, USA}
\affiliation{Center for Hybrid Nanostructures, Institute for Nanostructure and Solid State Physics, Universit\"at Hamburg, Luruper Chaussee 149, 22761 Hamburg, Germany}

\date{\today}

\begin{abstract}
We present an ultrafast single sub-micron particle detection method based on a half-bowtie coplanar waveguide. 
The method is capable of resolving the translocation of these particles at a bandwidth greater than $30\,\text{MHz}$. We compare
experimentally the simultaneous use of our radio-frequency technique with conventional DC based resistive pulse recordings and find that our method has a throughput that is enhanced by two orders of magnitude. The technique incorporates a microfluidic circuit and has potential to be employed for screening nano particles and biopolymers such as DNA at frequencies in excess of 1 GHz. 
\end{abstract}

\maketitle

One of the truly high-impact inventions in the field of bio-physics is the Coulter-counter \cite{h1953means, DeBlois1970}.
To this date it is found in applications world-wide and is being extended to counting sub-micron particles, strands of DNA, 
and even single molecules \cite{Bayley2000}. By now the topic forms a research field in its own right
and is part of many clinical applications due to its simplicity and high-throughput capabilities \cite{Venkatesan2011}.
The principle of operation is straightforward: a pore is embedded in a thin membrane, separating two reservoirs filled with electrolyte solution. A voltage across the membrane leads to an ionic current flow through the pore. Depending on the pore diameter this current ranges from nano-, for micron sized pores, to pico-amperes for nanopores. When a particle translocates through the pore the open pore current is altered by the presence of this particle because it partially blocks the ion flow and thus increases the resistivity of the pore. 

Although these setups have proven to work for a wide range of particle sizes down to the level where DNA sequencing becomes feasible~\cite{Laszlo2014} they still have one major drawback: their limitation in speed.
Especially, large access resistances and capacitances greatly limit the temporal resolution of the resulting data \cite{RosensteinIEEE}.  For the measurement of currents in the pA to nA range the use of operational amplifiers with large feedback resistances becomes mandatory. This resistance, coupled with the capacitive noise of the system, creates a poor signal-to-noise-ratio at high bandwidths (BWs) \cite{rosenstein}. Working towards
improving this, Fraikin~\cite{cleland_nano} and Rosenstein~\cite{rosenstein} have both made remarkable progress
by increasing the measurement BW and consequently the throughput by reducing the system's parasitic capacitance. Both report promising results with BWs around 500 kHz with the highest at 1~MHz. 

Nevertheless, the translocation velocity with which e.g. DNA slides through a nanopore (i.e. thousand nucleotides per millisecond  and more~\cite{Venkatesan2011}) is still much higher than the currently available speed of detection. In order to count or sequence at these high translocation speeds, it is compulsory to possess both, an extremely sensitive and extremely fast detector. That is if individual transitions are to be studied,
GHz-frequency BWs are required. 
One possible technique to meet this demand is the use of radio-frequency~(RF) circuits. 
Schoelkopf \textit{et al.} \cite{Schoelkopf} have shown that using a tank circuit and carrying out RF reflectometry helps to greatly enhance the sensitivity of their high impedance single electron transistor and increase  its  BW.

Our prior experiments have demonstrated that resolving the translocation of single molecules through an $\alpha$-hemolysin pore tracing the RF-response is indeed feasible~\cite{njp}. 
The necessary next step is to place such a setup in a planar fashion with an increased sensitivity, 
which we have recently reported on by using a coplanar waveguide (CPW) for tracing the formation of lipid bilayers in the frequency domain~\cite{firstPaper}. 
In this letter we bring these first results to fruition by an enhanced CPW tank circuit design, which 
resolves sub-micron particle translocations with BWs greater than $30\,\text{MHz}$.

\begin{figure}
\begin{center}
\centerline{\includegraphics[width=.45\textwidth]{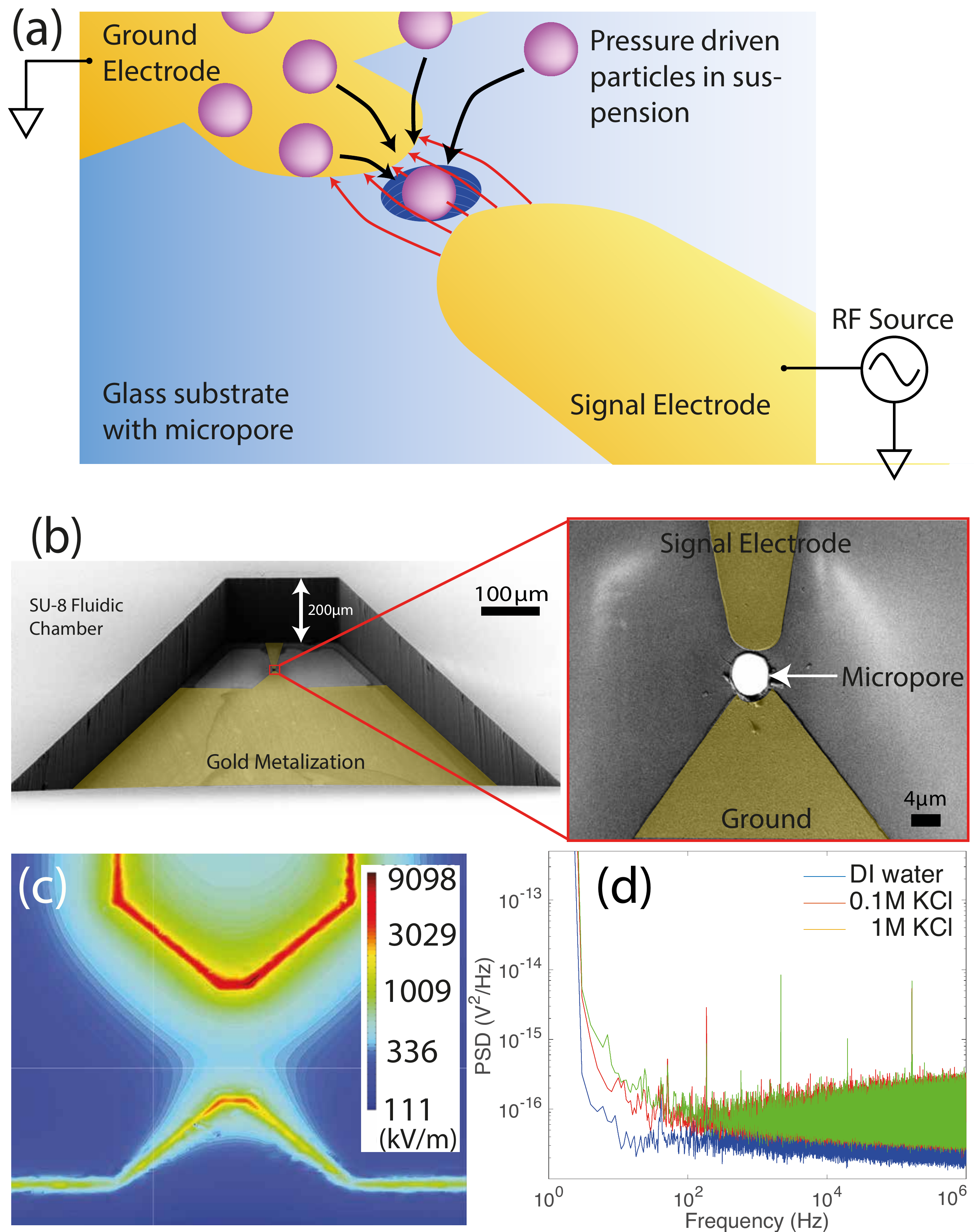}}
\caption{(a) Schematic of the chip design. The gold CPW is patterned on a microscope glass slide and the micropore is embedded within the open end of the CPW signal line. Test particles are introduced in suspension and a suction drives them through the pore. (b) Scanning electron microscopy images of the fluidic reservoir of the sensing region and zoomed-in view of a micropore embedded between the metal electrodes. (c) Electric fields in the sensing regions of the half bowtie CPW open circuits. The half bowtie has a tapering signal line and a small protrusion from the open circuit ground conductor. (d) Power spectral density of the reflected voltage signal.}
\label{fig:setupSchematic}
\end{center}
\end{figure}

A schematic view of the tank circuit can be seen in Fig.~\ref{fig:setupSchematic}:
the CPW is patterned on a $170\,\text{\textmu m}$ thick microscope glass slide by standard optical lithography. Gold with a thickness of about $110\,\text{nm}$ on top of a $5\,\text{nm}$ chrome adhesion layer is used as the CPW metalization. Within the sensing region a micropore is located, which allows for the vertical flow of particles. To characterize the chip we use polystyrene beads that translocate through the pore due to a negative pressure applied across the micropore. For measurement the chip is mounted in a custom made chip holder. When the sensing region, shaped as a bow-tie antenna, is flooded with electrolyte solution, DC measurements can be conducted by inserting Ag/AgCl  measurement electrodes in opposing sides of the chip holder (i.e. across the micropore such that current can flow through the pore when a voltage is applied).
To decrease the interference of the electrical field of the signal line with the electrolyte solution a $(100 - 200)\,\text{\textmu m}$ thick SU8 fluidic channel is patterned around the sensing region (see Fig. \ref{fig:setupSchematic} (b)) thus isolating the signal line from the solution. 
Micropores are drilled directly into the glass substrate using laser ablation via an ArF excimer laser \cite{Yu:09}. This method enables us to drill pores with tuneable pore diameters ranging from $100\,\text{nm}$ up to several micrometers (see inset in  Fig.~\ref{fig:setupSchematic}~(b)). For the experiments discussed in this paper the diameter of the laser drilled pore is about $2 \ \text{\textmu m}$.

At the bow tie apex, the end of the signal line is tapered to form a chamfered tip. A protrusion is introduced in the open circuit end of the ground plane, while the sides of the ground plane are moved out, thus increasing the gap between the sides of the signal and ground lines. 
These modifications enhance the field concentration at the tip of the signal line, while reducing the field on the sides. A direct consequence of which is an increase in sensitivity in the sensing region at the end of the signal line  (see Fig. \ref{fig:setupSchematic}(c)) and thus a substantial reduction in noise. 
 As a bead nears the vicinity of the pore, it is acted upon by the applied suction.

The sensor structure is designed to act as a tank circuit, which has a minimum in the reflected signal at a particular frequency. 
The technique relies on measuring the detuning of this tank circuit due to the particle translocating through the strong electrical field. This results in the amplitude modulation of the reflected voltage wave representing the position of the bead along the vertical axis.

\begin{figure}
\centerline{\includegraphics[width=.5\textwidth]{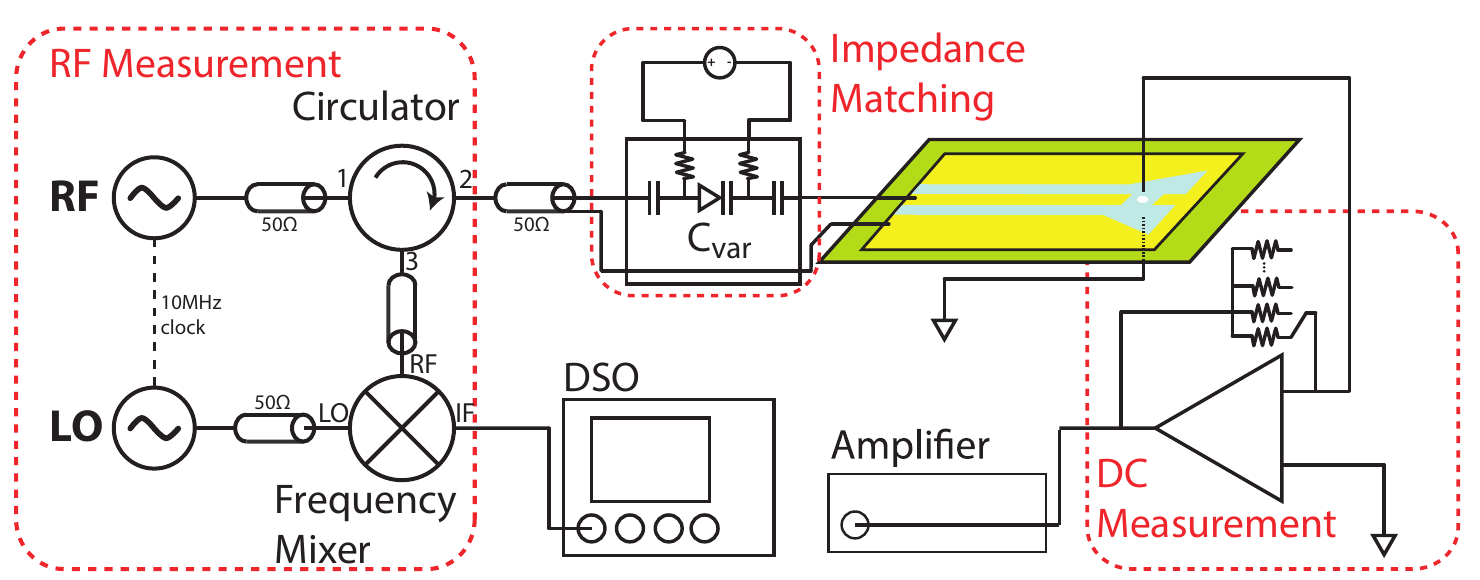}}
\caption{Schematic of the measurement setup. The RF signal is generated with a signal generator and fed into the sensor chip via an RF circulator. The sensor is connected to an impedance matching network comprising of a varactor diode, whose capacitance $C_\text{var}$ can be tuned using an external DC source. A mixer is used to mix the reflected RF signal with a local oscillator with a fixed frequency difference for downconversion of the signal. The response is measured with an oscilloscope (DSO). DC measurements are performed using a conventional  amplifier. }
\label{fig:Schematic}
\end{figure}


It can be shown that the best measurement performance can be achieved when the sensor impedance $Z_\text{L}$ equals the source impedance $Z_\text{0} = 50\,\Omega$ \cite{SteerRF}. The quality of this match is given by the reflection coefficient $S_{11}$ and is described as $S_{11} = \frac{Z_\text{L}-Z_0}{Z_\text{L}+Z_0} $. At the point of optimal impedance match the sensor impedance is real without any capacitive or inductive components. In Fig.~\ref{fig:setupSchematic} (d) the power spectrum density (PSD) of the amplitude of the reflected  RF wave is shown when the RF chip is impedance matched. The spectrum is acquired for different solutions with no salt (i.e. deionised water) as well as  $0.1\,\text{M KCl}$ and $1\,\text{M KCl}$ electrolyte. In contrast to conventional DC measurement strategies there is no high frequency noise component in the noise spectrum up to $1\,\text{MHz}$. 

Prior to carrying out time domain measurements, an impedance match was obtained for each device using a varactor-based tuneable matching network. Fig. \ref{fig:Schematic} shows a circuit diagram of the impedance matching circuit.
Varying the varactor diode capacitance $C_{\text{var}}$ by adjusting the voltage across the diode allows for setting $Z_{\text{L}} = Z_0$. At this point the power from the RF source is completely absorbed by the load and thus no reflection is measured. If the impedance is altered by the translocation of a bead through the pore it leads to a detuning of the circuit and thus to a reflection, which can be measured with great temporal resolution as will be discussed in the following.

\begin{figure}[thb!]
\begin{center}
\includegraphics[height=.8\textwidth]{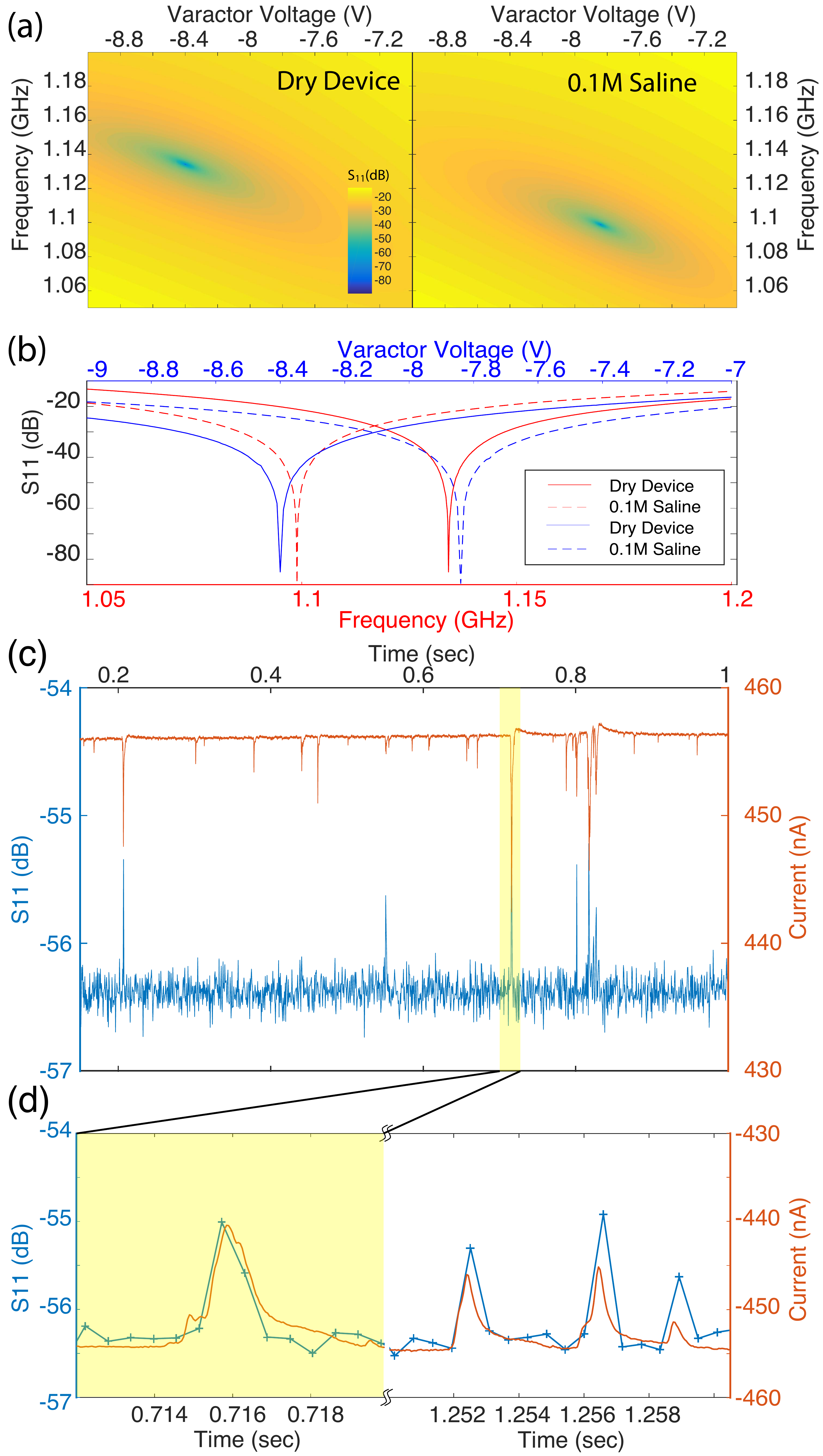}
\caption{(a) Direct measurement of $S_\text{11}$ vs. the frequency and varactor voltage to determine the point of minimal reflection, for a dry device and a device with electrolyte solution. The voltage and frequency step size is $10\,\text{mV}$ and $150\,\text{kHz}$, respectively. (b) Sections to the measurement seen in (a) for the best frequency and voltage response. (c) Measured translocation events of a suspension of polystyrene beads with diameters of  $500\;\text{nm}$ and $1\;\text{\textmu m}$, respectively. The red trace shows the DC measurement with a cutoff frequency of $3.4\;\text{kHz}$. The blue trace shows the corresponding RF trace that is acquired with an IFBW of $1\;\text{kHz}$. Alignment of the traces is exercised by determining the covariance for every time delay until a maximised covariance is found. (d) Zoomed in views into single events in the data partially shown in (c). The DC measurement (red trace) was inverted to show the correlation between the two measurements. }
\label{fig:VNA_Measurement}
\end{center}
\end{figure}

Fig.~\ref{fig:VNA_Measurement} (a) shows the results of impedance matching the sensor to the source before and after addition of saline in the sensing region. The measurements were acquired with a vector network analyser (VNA), which is connected to the impedance matching network, directly (see Fig.~\ref{fig:Schematic}). In this measurement for every varactor voltage the reflection $S_{11}$ was measured with respect to the RF signal. The data shows a characteristic minimum for each chip when it is dry and when it is filled with electrolyte solution. Fig.~\ref{fig:VNA_Measurement} (b) displays $S_{11}$ for the frequency and the voltage through the point of best match for a dry and a wet device. The voltage as well as the resonance frequency changes significantly thus giving a first impression on the sensitivity of our device.
The quality of the match was determined to be between $-60$ and $-90\,\text{dB}$ for all sensors tested. 

For a first measurement $S_{11}$ was sampled using a constant wave (CW) sweep at the resonance frequency of the chip using the VNA directly. This results in a reflection measurement vs. time for a chip with optimal impedance match.  The DC current through the pore was measured simultaneously. A suspension of polystyrene beads containing $500\,\text{nm}$ and $1\,\text{\textmu m}$ beads was used and the RF as well as DC response was measured independently over time. 
The result can be seen in Fig. \ref{fig:VNA_Measurement} (c): as the blue trace shows the time domain measurement of the reflected RF wave (i.e. $S_{\text{11}}$), the red trace shows the simultaneous DC measurement vs. time. A variation of $S_{11}$ to the best match point is typical as the point of best match is determined without an external pressure to hinder particle translocation during impedance matching. When time domain measurements are started a pressure is applied thus leading to a deviation from the optimal matching point. Nevertheless this does not hinder particle detection. For the measurement of  $S_{\text{11}}$ the signal is down-converted and bandpass filtered by the VNA. The BW of the bandpass filter is called intermediate frequency BW (IFBW). A reduction of the IFBW decreases the measured noise-floor and simultaneously decreases the measurement speed. This can be seen in Fig. \ref{fig:VNA_Measurement} (d). The points sampled by the VNA are shown in comparison to the DC measurement. Obviously every translocation event sampled by the VNA corresponds to only one to two data points. 
This reduction in measurement speed hinders the detection of each translocating particle. Nevertheless, the measurement gives a strong indication that the altered impedance due to the translocating beads create a measurable shift in the reflected wave. 

To increase the detection capability we implemented a heterodyne detection scheme (see  Fig.~\ref{fig:Schematic}). Similar to the measurement shown in Fig. \ref{fig:VNA_Measurement} (c) it enables us to measure the reflected RF response but at an IF and thus at a measurement BW and sampling rate that can be chosen freely. While the VNA measurement measures the reflection coefficient the heterodyne scheme will result in the sampling of the reflected RF wave, which is amplitude modulated due to the impedance mismatch caused by translocating particles. The reflected RF wave was sampled using a digital storage oscilloscope (DSO).

\begin{figure}
\begin{center}
\centerline{\includegraphics[width=.5\textwidth]{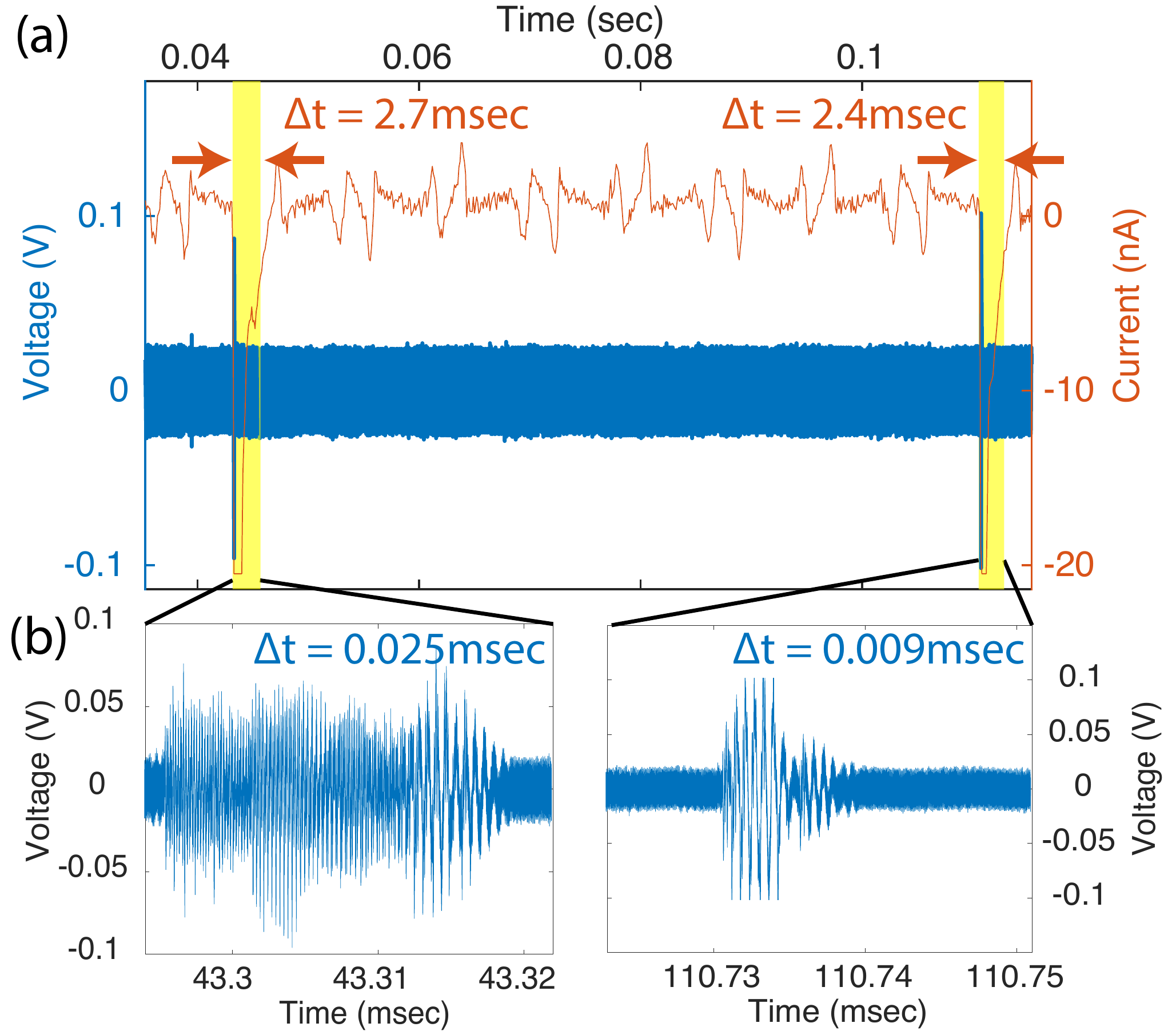}}
\caption{(a) Simultaneous RF (blue) and DC (red) measurement of $1\,\text{\textmu m}$ sized beads translocating through the half bowtie CPW sensor. The bead suspension used for those experiments were mono-dispersed in $145\,\text{mM}$ NaCl, $5~\text{mM}$ KCl, $1.8\,\text{mM}$ $\text{CaCl}_2$, $10\,\text{mM}$ HEPES  in DI water resulting in a $\text{pH} = 7.4$. (b) A zoomed in view of the reflected RF signals obtained from the beads passing through the sensor pore.}\label{fig:CompareRFDC}
\end{center}
\end{figure}

Results of experiments performed using a suspension of  $1\,\text{\textmu m}$ sized  beads on the device are displayed in Fig.~\ref{fig:CompareRFDC} (a). The RF trace illustrates at least two beads translocating through the sensor 67.4 ms apart, which is in accordance with the simultaneous DC measurement.  While the DC trace was sampled at 10 kHz and has a BW of 2.9 kHz, the RF trace has an effective BW in excess of $30\,\text{MHz}$, thus providing more detailed insight into the motion of the particle going through the pore. 

Fig.~\ref{fig:CompareRFDC}~(b) shows zoomed in views of the reflected RF signal obtained when the beads translocate through the pore. These amplitude modulated traces represent the interaction of the bead with the electrical field in the signal region delivering information about the motion of the particle as it approaches and traverses through the pore.  As the traces originate from a simultaneous DC and RF measurement another advantage of the measurement setup becomes apparent. The measured translocation time in the RF measurement is decreased by a factor 100. This is due to the fact that the length of the measured amplitude modulation is governed by the thickness of the gold metalization and not the length of the micropore. Consequently the potential throughput of our device is increased by approximately two orders of magnitude. It was shown by A. Fanget \textit{et al.} that it is possible to produce a NP, which is embedded between two platinum electrodes with thicknesses as small as $11\,\text{nm}$ on a silicon nitride membrane \cite{Fanget2014}. This is a crucial requirement for the application of our method to DNA screening without being dependent on thin insulating membranes as typically required in nanopore experiments \cite{Yanagi2015}.


The results demonstrate the degree of the CPW's sensitivity; a device capable of capturing the minuscule variations when beads are flowing through the pore. While the overall envelope of the traces are similar, the details of each trace are unique. These variations are caused due to the nature of each approaching particle (see Fig.~\ref{fig:CompareRFDC}~(b)). The peaks in the response suggest a particle in the pore and in the same plane as the metalization. The sensing system is a driven damped oscillator and the increasing oscillatory response leading up to the peak is a consequence of the time varying change in the impedance and hence in the resonant frequency of the sensor. This minute impedance change, caused by the translocating particle sets $Z_{L} \neq Z_0$ and results in an increase in $S_{11}$. 
 A relatively quick relaxation of the amplitude following the peak is by design and as expected due to the reduced sensitivity above and below the metalization plane. This indicates the passage of the particle and a return of the system to steady state with $Z_{L} = Z_0$.

The DC electrical recordings confirm that the CPW bowtie sensor can indeed be used to probe sub-micron particles at a BW greater than $30\,\text{MHz}$. The shape and magnitude of the response will depend on the material, size and shape of the particle under investigation. The duration of the response will depend on the transit time of the particle through the pore. If the device is operated at $100\,\text{MHz}$ BW it is possible to reliably detect sub-micron particles passing through the sensor at rates of over $10^8$ particles/sec. The device reliably detects the perturbation caused to the overall system impedance stemming from the dynamics of the particle in question.
Our setup allows the use of simultaneous DC measurements and the integration of microfluidics. As the measured translocation time of the particles is decreased by two orders of magnitude the potential throughput of the method is drastically increased. Furthermore, the measurement bandwidth at which the translocation can be acquired is increased by a factor of 30 in comparison to the best known method \cite{rosenstein}.  

We would like to thank the Deutsche Forschungsgemeinschaft (DFG) for funding via one
of their Excellence Centers (EXC-1024) and the Center for Ultrafast Imaging (CUI) at the University of
Hamburg. 

A.B. and P.V.G. contributed equally to this work.

\bibliography{refs}

\end{document}